\documentclass[journal]{IEEEtran}
\IEEEoverridecommandlockouts
\usepackage{array}
\usepackage{graphicx}
\usepackage{epsf}
\usepackage{float}
\usepackage{stfloats}
\usepackage[font=footnotesize]{caption}
\usepackage[font=footnotesize]{subcaption}
\usepackage{cite}
\usepackage{picinpar}
\usepackage{siunitx}

\usepackage{amsmath, nccmath}
\usepackage{url}
\usepackage{flushend}
\usepackage{colortbl}
\usepackage{soul}
\usepackage{multirow}
\usepackage{pifont}
\usepackage{color}
\usepackage{alltt}
\usepackage[hidelinks]{hyperref}
\usepackage{enumerate}
\usepackage{siunitx}
\usepackage{breakurl}
\usepackage{epstopdf}
\usepackage{pbox}

\usepackage{booktabs}
\usepackage{wrapfig}

\graphicspath{{../}}

\usepackage{placeins}
\usepackage{latexsym}
\usepackage{amssymb}
\usepackage{xifthen}
\usepackage{balance}
\usepackage{multicol}

\usepackage{multirow}
\usepackage{lipsum}
\usepackage{scalerel}

\usepackage{authblk}

\usepackage{nicefrac}

\definecolor{gray}{RGB}{0,0,0}

\ifCLASSINFOpdf

\else
 
\fi

\DeclareMathOperator{\eig}{eig}

\begin{document}

\title{Internal-Model-Control-Based Virtual Admittance Emulation for Enhanced Grid-Forming Performance}
\author{
	\vskip 1em
	
	Ru\v{z}ica Cvetanovi\'{c},
	Lazar Stojanovi\'{c} \emph{Student Member, IEEE}
    Paolo Sbabo, \emph{Student Member, IEEE}
	Paolo Mattavelli, \emph{Fellow, IEEE},
	Massimo Bongiorno, \emph{Senior Member, IEEE}
\thanks{
This work has been submitted to IEEE  for possible publication and is currently under peer review. Copyright may be transferred without notice, after which this version may no longer be accessible. \\
%
%
R. Cvetanovi\'{c} is with Innovation Center of the School of Electrical Engineering in Belgrade, 11000 Belgrade, Serbia (e-mail: ruzica.cvetanovic@ic.etf.bg.ac.rs). \\
L. Stojanovi\'{c} and P. Mattavelli are with Department of Management and Engineering, University of Padova, 36100 Vicenza, Italy (e-mails: lazar.stojanovic@phd.unipd.it and paolo.mattavelli@unipd.it). \\
P. Sbabo is with Department of Information Engineering, University of Padova, 35131 Padova, Italy (e-mail: paolo.sbabo@phd.unipd.it).\\
M. Bongiorno is with Department of Electrical Engineering, Chalmers University of Technology, Gothenburg, Sweden (e-mail: massimo.bongiorno@chalmers.se).
} 

}

\maketitle

\begin{abstract}
The large-scale integration of renewable energy into electrical power systems places a growing reliance on grid-forming converters (GFMs) to overcome critical stability and control hurdles. By emulating a slowly varying voltage source behind a tunable resistive–inductive impedance, GFMs enable precise setpoint tracking and vital grid support. In this regard, virtual admittance (VA)-based inner-loop control provides exceptional flexibility in shaping the converter's small-signal immittance at connection terminals, allowing for advanced GFM functionalities. This article addresses a novel internal-model-control-based VA approach, which inherently accounts for the effects of the inner current-control loop and the voltage feedforward. Unlike the state-of-the-art VA control, the proposed VA control ensures that, at frequencies outside the influence of outer loops, the converter's terminal admittance closely matches the target admittance, which is shown to be relevant for stability. It is revealed how, with the state-of-the art VA control, a small-signal instability in the harmonic range may arise if a low-pass filter with a high cut-off frequency is used for the voltage feedforward. However, the proposed VA control ensures stable operation regardless of this cut-off frequency. The methodology is experimentally validated using a laboratory prototype that features three-phase voltage-source converters.

\end{abstract}


\IEEEpeerreviewmaketitle
\vspace{-1em}
\section{Introduction}

\IEEEPARstart{G}{rid}-forming converters (GFMs) have proven effective for maintaining stability and robustness in low-inertia, converter-dominated power systems \cite{rossoGridFormingConvertersControl2021,tozakModelingControlGrid2024,imgartExternalInertiaEmulation2024,matevosyanGridFormingInvertersAre2019,entso-eHighPenetrationPower2020,zhongSynchronvertersInvertersThat2011}. Typically, GFMs aim to emulate a slowly varying internal voltage source (IVS) behind a tunable resistive–inductive impedance \cite{narulaPhD,kamalinejadImpactControlParameters2025}. For this, a multi-loop control architecture is commonly employed, where the slower outer loops regulate active and reactive power (or terminal voltage magnitude) by determining the synchronization angle and IVS magnitude that the faster inner loops should achieve \cite{rossoGridFormingConvertersControl2021,tozakModelingControlGrid2024,vattakkuniComparativeAssessmentTypical2023,zhaoExploringDampingEffect2025,narulaPhD,liuUnifiedVoltageControl2024,pengImprovedReactanceControl2026}. In this context, virtual admittance (VA) control has proven effective for achieving advanced GFM functionalities, by improving the flexibility in shaping the converter's small-signal immittance at connection terminals and partially decoupling the IVS dynamics from the grid voltage dynamics \cite{imgartDecoupledPQGridForming2026,anant_GFM_comparison,eggersAccuracyStabilityAssessment2026,zhaoAnalysisActiveDamping2026}. To this end, a variety of VA control methods have been proposed \cite{huangImpactVirtualAdmittance2021,leonGridFormingControllerBased2023,rodriguezControlGridconnectedPower2013,eggersAccuracyStabilityAssessment2026}. While these inner-loop control approaches differ in their specific implementation, they share the common objective of shaping the converter’s terminal immittance to match the desired reference.

Concurrently, as grid-code requirements for GFMs continue to advance, evaluating converter's behavior in the frequency domain has become increasingly important for demonstrating compliance \cite{GBGF2023,AEMO2024,fingrid}. Frequency-domain analysis offers a practical framework not only for verifying adherence to these requirements but also for characterizing essential GFM properties, particularly those related to stability \cite{entso-eHighPenetrationPower2020,cardozoPromisesChallengesGrid2024}. Most existing VA control strategies \cite{huangImpactVirtualAdmittance2021,leonGridFormingControllerBased2023,rodriguezControlGridconnectedPower2013,eggersAccuracyStabilityAssessment2026} have been developed primarily to achieve accurate immittance tracking at low frequencies, thus providing a solid basis for the outer loop design. Nevertheless, to further exploit GFM capabilities, it is of interest to extend the frequency range where accurate immittance emulation is achieved. 

Precise admittance shaping in the harmonic range can improve a GFM's ability to support the grid and act as a sink to counter harmonics in the grid voltage \cite{anant_GFM_comparison,entso-eHighPenetrationPower2020}. Moreover, by ensuring adherence to the resistive-inductive behavior, GFM's passivity properties can be improved, thereby preventing the associated destabilizing interactions \cite{wangHarmonicStabilityPower2019,entso-eHighPenetrationPower2020,avdiajPassivityBasedDesignMethodology2025,harneforsInputAdmittanceCalculationShaping2007,harneforsPassivityBasedControllerDesign2015,eggersAccuracyStabilityAssessment2026,huangImpactVirtualAdmittance2021}. Accordingly, ensuring target immittance emulation also in the harmonic range can be considered vital for mitigating such interactions, meeting evolving standards, and enabling advanced GFM functionalities \cite{entso-eHighPenetrationPower2020,cardozoPromisesChallengesGrid2024}. To the best of the authors' knowledge, VA emulation accuracy over the wide frequency range, and its influence on GFM stability in the harmonic range, have not been addressed so far. 

To address these gaps, this article, building on \cite{ipec2026,caldognettoImpedanceSynthesisInverter2016,caldognettoPowerElectronicsBased2017}, illustrates how the conventional VA approach \cite{rodriguezControlGridconnectedPower2013,huangImpactVirtualAdmittance2021}, in which the VA controller is simply assigned the transfer function of the target admittance, fails to achieve the target frequency response at higher frequencies. It is explained how this limitation arises from the effect of the current control loop and the voltage feedforward, both of which are neglected in the conventional VA design. Next, the impact of voltage feedforward filtering on stability is addressed. It is revealed how, with the conventional VA control, an instability in the harmonic range may arise in case a low-pass filter with a high cut-off frequency is used for the voltage feedforward. It is demonstrated how this instability can be successfully prevented by using a novel internal-model-control-based VA approach \cite{ipec2026}. By accounting for the current control loop dynamics and impact of feedforward, the novel VA control ensures that the GFM's admittance matches the target in a wide frequency range, which is demonstrated as advantageous from the stability point-of-view. The impact of admittance shaping on stability is analyzed by comparing the passivity properties of the GFM's admittance with different VA design approaches, illustrating assets of the novel VA control. Its positive impact on stability was demonstrated in various grid-connected scenarios. The proposed methodology is experimentally validated in both the time and frequency domains, using a laboratory prototype comprising three-phase voltage-source GFMs.

The remainder of this article is organized as follows. Section II introduces the considered grid-forming control system. Section III addresses conventional VA control and illustrates its negative impact on admittance passivity and system stability. Section IV addresses the novel, internal model-control based VA approach and demonstrates its positive impact on admittance passivity and system stability. Section V presents experimental validation. Section VI concludes the article.

\section{Considered Grid-Forming Control System}

\begin{figure*}[!t]
	\centering
	\includegraphics[width=\linewidth]{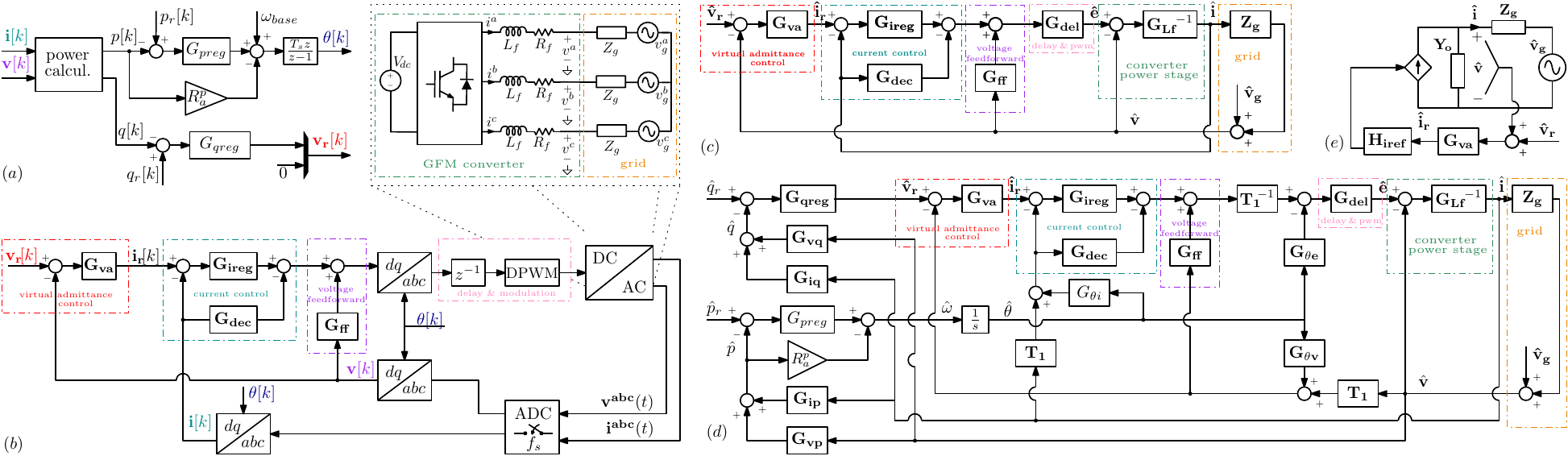}
	\caption{Block diagram of the considered GFM control: (a) outer-loops; (b) inner-loops. Small-signal $dq$ frame representation: (c) inner-loops; (e) outer- and inner-loops. (d) equivalent representation of (c), which highlights impact of current reference perturbation $\mathbf{\hat{i}_r}$ on GFM's closed-loop admittance $\mathbf{Y}$.}
	\label{fig:bd}
\end{figure*}

To demonstrate the proposed methodology, Fig. \ref{fig:bd} illustrates a representative grid-forming control architecture. Here, outer loops regulate active and reactive power to determine the synchronization angle $\theta$ and virtual back-electromotive force (EMF) reference $\mathbf{v_r}$ \cite{narulaPhD}. The primary focus of this article is the synthesis of a precise VA, establishing a generic inner-loop foundation that can be integrated with different outer-loop control schemes.

The considered VA-based inner loop architecture is shown in Fig. \ref{fig:bd}(b), where the inner loops are implemented in the synchronous rotating ($dq$) frame; however, the principles presented hereafter are equally applicable when either the current control or the VA control is implemented in the stationary ($\alpha\beta$) frame. As shown in Fig. \ref{fig:bd}(b), the difference between the target EMF $\mathbf{v_r}$ and the voltage at the GFM's connection point $\mathbf{v}$ is processed by the VA controller $\mathbf{G_\mathrm{va}}$, to find the current reference $\mathbf{i_r}$. The difference between $\mathbf{i_r}$ and the current $\mathbf{i}$ is processed by the current-controller $\mathbf{G_\mathrm{ireg}}$. The $dq$ axis current decoupling action is implemented through $\mathbf{G_\mathrm{dec}}$. Finally, the voltage feedforward term is added, which involves processing the measured voltage $\mathbf{v}$ through a low-pass filter $\mathbf{G_\mathrm{ff}}$.

\subsection{Small-signal $s$-domain model of inner control loops}
Small-signal $dq$ frame $s$-domain representation of the considered inner-loop control system architecture from Fig. \ref{fig:bd}(b) is shown in Fig. \ref{fig:bd}(c). The vectors $\mathbf{\hat{i}} = [\hat{i}^d, \hat{i}^q]$, $\mathbf{\hat{v}} = [\hat{v}^d, \hat{v}^q]$, $\mathbf{\hat{i}_r} = [\hat{i}_r^d, \hat{i}_r^q]$, $\mathbf{\hat{v}_r} = [\hat{v}_r^d, \hat{v}_r^q]$, $\mathbf{\hat{e}} = [\hat{e}^d, \hat{e}^q]$  represent the small-signal $dq$ components of the GFM's inductor current, terminal voltage, current reference, target EMF, and the switched-node voltage, respectively. As for the transfer function matrices, $\mathbf{Z_g}(s)$ represents the grid's impedance, while
\begin{equation}
	\mathbf{G_{Lf}}(s) = 	\begin{bmatrix}
		sL_f + R_f & -\omega_{g}L_f \\
		\omega_{g}L_f & sL_f + R_f 
	\end{bmatrix}
	\label{eq:GLf}
\end{equation}
models the converter's inductive output filter. In \eqref{eq:GLf}, $\omega_{g}$ is the steady-state value of the angular frequency of the ac-side quantities, which, under nominal conditions is equal to the base frequency $\omega_\mathrm{base} = 2 \pi f_\mathrm{base}$. Next, $L_f = L_f^{pu}Z_\mathrm{base}/\omega_\mathrm{base}$, $R_f = R_f^{pu}Z_\mathrm{base}$, $L_f^{pu}$ and $R_f^{pu}$ are converter's output filter inductance and series resistance absolute and per unit values, respectively, $Z_\mathrm{base} = 1/Y_\mathrm{base}$ is the base impedance. Further,
\begin{equation}
	\mathbf{G_\mathrm{ff}}(s) = 	\begin{bmatrix}
		\frac{\omega_\mathrm{ff}}{s+\omega_\mathrm{ff}} & 0 \\
		0 & \frac{\omega_\mathrm{ff}}{s+\omega_\mathrm{ff}}
	\end{bmatrix}
	\label{eq:Gff}
\end{equation}
models low-pass filter used for voltage feedforward, where $\omega_\mathrm{ff} = 2\pi f_\mathrm{ff}$ and $f_\mathrm{ff}$ is the filter's cut-off frequency. Next,
\begin{equation}
	\mathbf{G_\mathrm{\!del}}(s) = 	\begin{bmatrix}
		\cos({\omega_{g}\tau_d}) & \sin({\omega_{g}\tau_d}) \\
		-\sin({\omega_{g}\tau_d}) & \cos({\omega_{g}\tau_d})
	\end{bmatrix}e^{-s\tau_d}
	\label{eq:Gdel}
\end{equation}
is the $dq$ frame representation of the delay $\tau_d$, which in typical digital control implementation is present due to computation time of digital control algorithm and due to the digital pulse-width modulation (DPWM). The transfer function matrix of the current controller is given by
\begin{equation}
	\mathbf{G_\mathrm{ireg}}(s) = 	\begin{bmatrix}
		k_p + \frac{k_i}{s} & 0 \\
		0 & k_p + \frac{k_i}{s}
	\end{bmatrix}
	\label{eq:Gireg}
\end{equation}
where $k_p = \omega_\mathrm{ireg}L_f$, $k_i = \omega_\mathrm{ireg}R_f$ are proportional and integral gain of the current controller, which are tuned to achieve the desired bandwidth $\omega_\mathrm{ireg} = 2 \pi f_\mathrm{ireg}$ and phase margin of the current control loop. The current-controller's $dq$ axis decoupling term is given by
\begin{equation}
	\mathbf{G_\mathrm{dec}} = 	\begin{bmatrix}
		0 & \omega_g L_f \\
		-\omega_g L_f & 0
	\end{bmatrix}.
	\label{eq:Gdec}
\end{equation}
VA controller transfer function matrix is
\begin{equation}
	\mathbf{G_\mathrm{va}}(s) =\begin{bmatrix}
		G_\mathrm{va}^{\mathrm{dd}}(s) & G_\mathrm{va}^{\mathrm{dq}}(s) \\
		G_\mathrm{va}^{\mathrm{qd}}(s) & G_\mathrm{va}^{\mathrm{qq}}(s)
	\end{bmatrix}
	\label{eq:Gvreg_dq}
\end{equation}
Its specific structures are discussed in the next two sections, where performance of VA controllers is characterized by evaluating GFM admittance $\mathbf{Y}(s)$ at the point of connection
\begin{equation}
	\mathbf{\hat{i}}(s) \!= -\mathbf{Y}(s) \mathbf{\hat{v}}(s) = -
	\begin{bmatrix}
		Y_{\mathrm{dd}}(s) & Y_{\mathrm{dq}}(s) \\
		Y_{\mathrm{qd}}(s) & Y_{\mathrm{qq}}(s) 
	\end{bmatrix}\mathbf{\hat{v}}(s)
	\label{eq:ydef}
\end{equation}
According to \eqref{eq:ydef} and Fig. \ref{fig:bd}(c), the expression for $\mathbf{Y}(s)$ with only inner loops ($\hat{\theta} = 0$, $\mathbf{\hat{v}_r} = \mathbf{0}$ in Fig. \ref{fig:bd}(d)) is given by
\begin{multline}
		\mathbf{Y_\mathrm{\!inner}}(s) = \big(\mathbf{G_{Lf}}(s) + \mathbf{G_\mathrm{\!del}}(s)\left(\mathbf{G_\mathrm{dec}} +\mathbf{G_\mathrm{ireg}}(s)\right)\big)^{-1} \\  \big(\mathbf{I} + \quad \mathbf{G_\mathrm{\!del}}(s)\mathbf{G_\mathrm{ireg}}(s)\mathbf{G_\mathrm{va}}(s)-\mathbf{G_\mathrm{\!del}(s)}\mathbf{G_\mathrm{ff}}(s)\big).
	\label{eq:y}
\end{multline}
where $\mathbf{I}$ is the identity matrix. Also, \eqref{eq:y}, can be written as
\begin{equation}
    \mathbf{Y_\mathrm{\!inner}}(s) = \mathbf{Y_o}(s) + \mathbf{H_{iref}}(s)\mathbf{G_\mathrm{va}}(s).
    \label{eq:y_yo_hiref}
\end{equation}
where 
\begin{equation}
	\mathbf{Y_o}(s) = \mathbf{G}(s)^{-1} \big(\mathbf{I} -\mathbf{G_\mathrm{\!del}}(s)\mathbf{G_\mathrm{ff}}(s)\big)
	\label{eq:yo}
\end{equation}
is the converter's admittance in case current reference perturbation is nulled ($\mathbf{\hat{i}_{r}}(s) = \mathbf{0}$), 
\begin{equation}
	\mathbf{H_{iref}}(s) = \mathbf{G}(s)^{-1}\mathbf{G_\mathrm{\!del}}(s)\mathbf{G_\mathrm{ireg}}(s)
	\label{eq:hiref}
\end{equation}
is the current reference tracking transfer function matrix, and
\begin{equation}
    \mathbf{G}(s) = \mathbf{G_{Lf}}(s) + \mathbf{G_\mathrm{\!del}}(s)\left(\mathbf{G_\mathrm{dec}} +\mathbf{G_\mathrm{ireg}}(s)\right)(s).
\end{equation}
Fig. \ref{fig:bd}(e) indicates how current loop dynamics impacts $\mathbf{Y_\mathrm{\!inner}}$ via both $\mathbf{H_{iref}}$ and $\mathbf{Y_o}$, while voltage-feedforward impacts $\mathbf{Y_\mathrm{\!inner}}$ only via $\mathbf{Y_o}$. Thus, as elaborated in Section \ref{sec:ap2}, to achieve accurate admittance emulation in a wide frequency range (outside the influence of outer loops), design of $\mathbf{G_\mathrm{va}}$ must account for the impact of both $\mathbf{Y_o}$ and $\mathbf{H_{iref}}$.

\subsection{Small-signal $s$-domain model with outer control loops}

Fig. \ref{fig:bd}(d) shows small-signal $dq$ frame $s$-domain representation of the considered control system architecture from Fig. \ref{fig:bd}(a)-(b), which includes both inner- and outer-loops. The model is based on \cite{guoImpedanceAnalysisStabilization2021}. While the full transfer function/matrices and vector/matrix gain expressions are omitted for brevity, their physical meaning is outlined below. $G_\mathrm{preg}(s)$ models proportional-integral active power controller; $\mathbf{G_\mathrm{\mathbf{qreg}}}(s)$ models integral reactive power controller; the steady-state operating-point dependent gains $\mathbf{G_{vp}}$, $\mathbf{G_{ip}}$, $\mathbf{G_{vq}}$, $\mathbf{G_{iq}}$ model the power dependence on voltage and current, while $\mathbf{T_1}$, $\mathbf{G_{\theta i}}$, $\mathbf{G_{\theta v}}$, $\mathbf{G_{\theta e}}$ model $dq$ frame synchronization effect. Acc. to \eqref{eq:ydef} and Fig. \ref{fig:bd}(d) ($\hat{p}_r = 0$, $\hat{q}_r = 0$) the GFM's admittance with both inner and outer loops is given by
\begin{multline}
\mathbf{Y_\mathrm{\!outer}}(s)\!=\!\left( \mathbf{G}_\mathrm{\!del}\mathbf{T}_1^{-1}\mathbf{M}_1 \! - \! \mathbf{G}_{\!L_f} \! - \!\mathbf{G}_\mathrm{\!del}\mathbf{M}_2\mathbf{G}_\mathrm{preg}^\Sigma\mathbf{G}_{ip} \right)^{\!-1} \\
\cdot \left( \mathbf{G}_\mathrm{\!del}\mathbf{T}_1^{-1}\mathbf{M}_3 - \mathbf{I} + \mathbf{G}_\mathrm{\!del}\mathbf{M}_2\mathbf{G}_\mathrm{preg}^{\Sigma}\mathbf{G}_{vp} \right)
\label{eq:y_outer}
\end{multline}
where
\begin{gather*}
\mathbf{M}(s) = -(\mathbf{G}_\mathrm{dec}+\mathbf{G}_\mathrm{ireg})\mathbf{T}_1 - \mathbf{G}_\mathrm{ireg}\mathbf{G}_\mathrm{va}\mathbf{G}_\mathrm{qreg}\mathbf{G}_{iq} \\
\begin{aligned}[b]
\mathbf{M}_2(s) = -\mathbf{G}_{\theta e} + \mathbf{T}_1^{-1} \big( & \mathbf{G}_\mathrm{ff}\mathbf{G}_{\theta v} - (\mathbf{G}_\mathrm{dec}+\mathbf{G}_\mathrm{ireg})\mathbf{G}_{\theta i} \\
& - \mathbf{G}_\mathrm{ireg}\mathbf{G}_\mathrm{va}\mathbf{G}_{\theta v} \big)
\end{aligned} \\
\mathbf{M}_3(s) = \mathbf{G}_\mathrm{ff}\mathbf{T}_1 - \mathbf{G}_\mathrm{ireg}\mathbf{G}_\mathrm{va}(\mathbf{T}_1 + \mathbf{G}_\mathrm{qreg}\mathbf{G}_{vq}).
\end{gather*}
and $G_{preg}^\Sigma = (G_{preg} + R_a^p)/s$. Note that substitution of $\mathbf{G_\mathrm{preg}} = \mathbf{G_\mathrm{qreg}} = \mathbf{G_{vp}} = \mathbf{G_{ip}} = \mathbf{G_{vq}} = \mathbf{G_{iq}} = \mathbf{G_{\theta i}} = \mathbf{G_{\theta v}} = \mathbf{G_{\theta e}} = \mathbf{0}$, and $\mathbf{T_1} = \mathbf{I}$ in \eqref{eq:y_outer} yields \eqref{eq:y}, i.e., admittance when outer loops are neglected.

\begin{table}[]
	\caption*{Table I: Parameters of the converter used for the case study.}
	\centering
	\begin{tabular}{lccc}
		\midrule\midrule
		Parameter & label & value & unit\\
		\midrule
		Base power & $S_\mathrm{base}$ & $100$ &  MW  \\     
		Base voltage & $V_\mathrm{base}$ & $400$ &  kV  \\
		Base (grid) frequency & $f_\mathrm{base}$ & $50$ &  Hz  \\
		Filter inductance & $L_f^{pu}$ & $0.15$ & p.u. \\
		Filter resistance & $R_f^{pu}$ & $0.015$ & p.u. \\
		Sampling frequency & $f_{s} = \frac{1}{T_s}$ & $4$ & kHz \\
		Virtual inductance & $L_v^{pu}$ & $0.3$ & p.u. \\
		Virtual resistance & $R_v^{pu}$ & $0.15$ & p.u. \\
		Current loop bandwidth & $f_\mathrm{ireg}$ & $200$ & Hz \\
		Feedforward cut-off frequency & $f_{\!f\!f}$ & $\{50, 1500\}$ & Hz \\
		Control loop delay & $\tau_{d}$ & $[0,375]$ & $\mu$s \\
        Outer (PQ) loop bandwidth & $f_{\mathrm{pqreg}}$ & $5$ & Hz \\
		\midrule\midrule
		
	\end{tabular}
\end{table}

\section{Conventional Virtual Admittance Control}\label{sec:ap1}

A commonly adopted approach  to implement a VA controller involves \cite{rodriguezControlGridconnectedPower2013}
\begin{equation}
	\mathbf{G_\mathrm{va}}(s) = \mathbf{Y_{v}}(s)
	\label{eq:Gvreg_conv}
\end{equation}
where
\begin{equation}
	\mathbf{Y_{v}}(s) =\begin{bmatrix}
		sL_v + R_v & -\omega_{g}L_v \\
		\omega_{g}L_v & sL_v + R_v 
	\end{bmatrix}^{-1}
	\label{eq:yv}
\end{equation}
is the target admittance to achieve by inner loops. Based on \eqref{eq:Gvreg_dq}, \eqref{eq:Gvreg_conv} and \eqref{eq:yv}
\begin{equation}
    G_\mathrm{va}^{\mathrm{dd}}(s) = G_\mathrm{va}^{\mathrm{qq}}(s) = \frac{sL_v + R_v}{(sL_v + R_v)^2 + (\omega_{g}L_v)^2}
    \label{eq:Gva_conv_dd_qq}
\end{equation}
\begin{equation}
    G_\mathrm{va}^{\mathrm{dq}}(s) = -G_\mathrm{va}^{\mathrm{qd}}(s) = \frac{\omega_{g}L_v}{(sL_v + R_v)^2 + (\omega_{g}L_v)^2}
     \label{eq:Gva_conv_dq_qd}
\end{equation}
Thus, according to \eqref{eq:Gva_conv_dd_qq} and \eqref{eq:Gva_conv_dq_qd}, $s$-domain implementation of the conventional VA controller in the $dq$ frame involves four second order transfer functions. For digital implementation, any of the standard discretization methods can be used. In this article, Tustin method is used.

\begin{figure*}[!t]
	\centering
	\includegraphics[width=\linewidth,trim={0 0 0.5cm 0},clip]{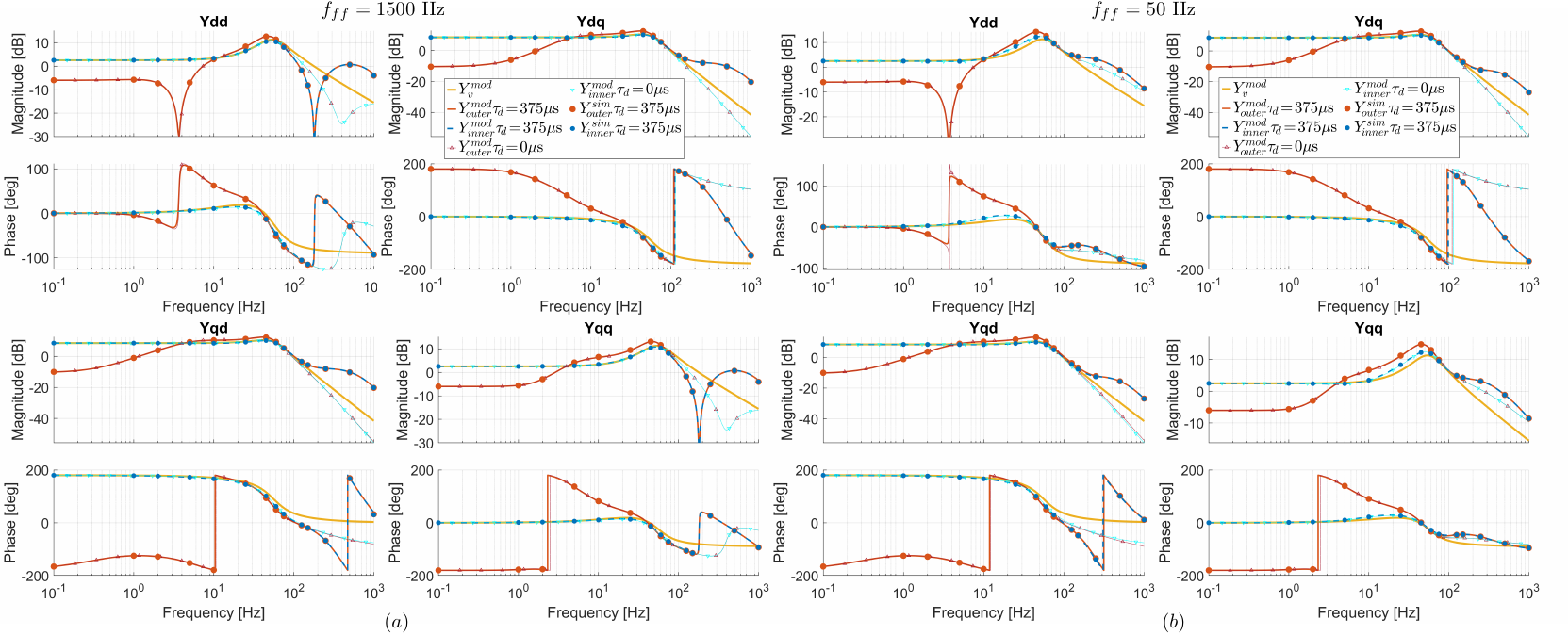}
	\caption{Normalized admittance frequency responses of the converter from Fig. \ref{fig:bd} and Table I when conventional VA approach is used: (a) $f_\mathrm{ff} = 1500$ Hz; (b) $f_\mathrm{ff} = 50$ Hz. Comparison between results from simulation (dot markers) and analytical model (lines) with/out outer loops ($P_{r} = 0.5$ pu, $Q_{r} = 0.3$ pu).}
	\label{fig:y_ap1}
\end{figure*}

\begin{figure*}[!t]
	\centering
	\includegraphics[width=\linewidth,trim={0 0 0.05cm 0},clip]{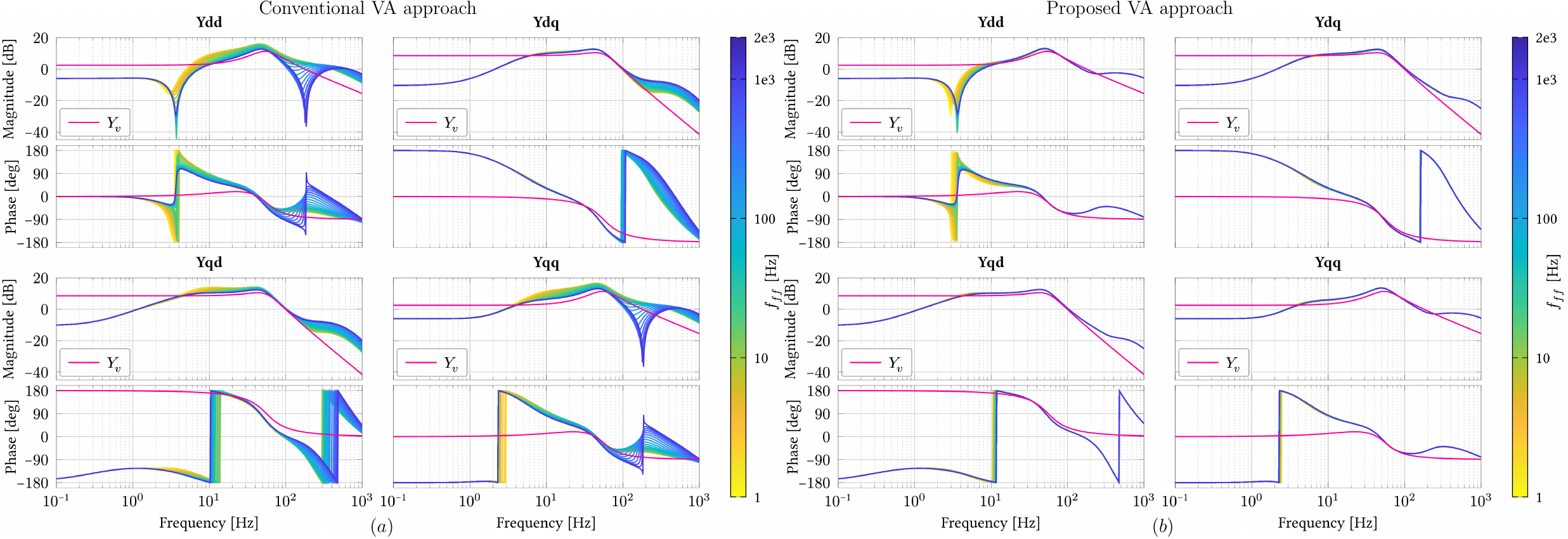}
	\caption{Normalized admittance frequency responses obtained from the analytical model with outer loops \eqref{eq:y_outer} for the converter from Table I ($\tau_d = 375$ $\mu$s) and Fig. \ref{fig:bd} with (a) conventional; (b) proposed VA approach.}
	\label{fig:sweep}
\end{figure*}

\subsection{Frequency Domain Accuracy Characterization}

To examine the performance of this conventional VA approach, GFM from Fig. \ref{fig:bd} and Table I is considered. Fig. \ref{fig:y_ap1}(a) and (b) show the GFM's admittance when $f_\mathrm{ff} = 1500$ Hz and $f_\mathrm{ff} = 50$ Hz, respectively. The results from  MATLAB/Simulink switching-period-averaged simulation with discrete-time control implementation and those from the analytical model with \eqref{eq:y_outer} and without outer loops \eqref{eq:y} are compered for $\tau_d = 0$ and $\tau_d = 375$ $\mu$s. Frequency responses of the target admittance $\mathbf{Y_v}$ (from \eqref{eq:yv}) are also shown. Several remarks can be made from Fig. \ref{fig:y_ap1}.

First, the results from simulation and analytical model are consistent, which validates the latter. Second, with outer loops, the low-frequency asymptotes are determined by the operating point, whereas with inner loops only, they are determined by the target virtual admittance. At high frequencies, above approximately $100$ Hz in the considered example, the results with and without outer loops overlap. The specific frequency value above which the outer loop impact becomes negligible is mostly determined by the outer-loop bandwidth, but also affected by the inner-loop design. Typically, the outer-loop impact is negligible in the harmonic range, which justifies considering only inner-loops for the analysis in this range. In the harmonic range, which is the focus of this article, the following can be concluded from Fig. \ref{fig:y_ap1}. Even without any delays ($\tau_d = 0$), the converter's admittance $\mathbf{Y}$ significantly deviates from the target admittance $\mathbf{Y_v}$. This is mainly due to the impact of the current control loop and the voltage feedforward. Digital delays also contribute to the loss of admittance emulation accuracy. Nevertheless, the focus of this study remains on the effects of the current controller and the voltage feedforward.

To further illustrate the impact of the voltage feedforward with the conventional VA approach, Fig. \ref{fig:sweep}(a) is provided. There, the converter's admittance frequency responses obtained from the analytical model with outer loops \eqref{eq:y_outer} are plotted for $25$ different values of the feedforward low-pass filter's cut-off frequency $f_\mathrm{ff}$, logarithmically spaced across the range [$1$,$2000$] Hz. The value of $f_\mathrm{ff}$ significantly impacts admittance frequency responses. For lower values of $f_\mathrm{ff}$, this impact extends down into the sub-synchronous range. Although most pronounced for high values of $f_\mathrm{ff}$, for all considered $f_\mathrm{ff}$ values, the GFM's admittance deviates substantially from the target profile in the harmonic range. As elaborated below, this can severely compromise system stability.

\subsection{Admittance Passivity and System Stability}
\begin{figure}[!t]
	\centering
	\includegraphics[width=\linewidth]{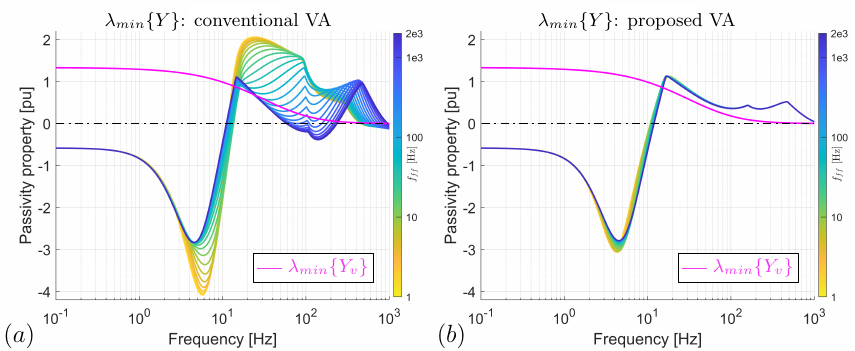}
	\caption{Passivity properties of the admittances from Fig. \ref{fig:sweep}.}
	\label{fig:sweep_2}
\end{figure}

To illustrate how, depending on the value of $f_\mathrm{ff}$, the conventional VA approach may compromise system stability, impact of $f_\mathrm{ff}$ on the GFM's admittance passivity property 
\begin{equation}
	\lambda_\mathrm{min}(j\omega) = \min{ \left( \eig \left( \frac{\mathbf{Y}(j\omega)+\mathbf{Y}^H(j\omega)}{2} \right) \right)},
	\label{eq:lamda}
\end{equation}
is examined, where $\eig$ returns the eigenvalues of its argument and $H$ is the Hermitian (conjugate transpose) operator. Fig. \ref{fig:sweep_2}(a) shows the passivity properties of the admittances from Fig. \ref{fig:sweep}(a). Although lower $f_\mathrm{ff}$ values are advantageous for passivity properties in the harmonic-range, the negative dip of $\lambda_\mathrm{min}$ in the sub-synchronous range gets worse when $f_\mathrm{ff}$ is low. For higher values of $f_\mathrm{ff}$, $\lambda_\mathrm{min}$ becomes negative around $100$-$200$ Hz (in $dq$). In presence of poorly damped grid anti-resonances, this may be detrimental for system stability.

\begin{figure}[!t]
	\centering
	\includegraphics[width=0.9\linewidth]{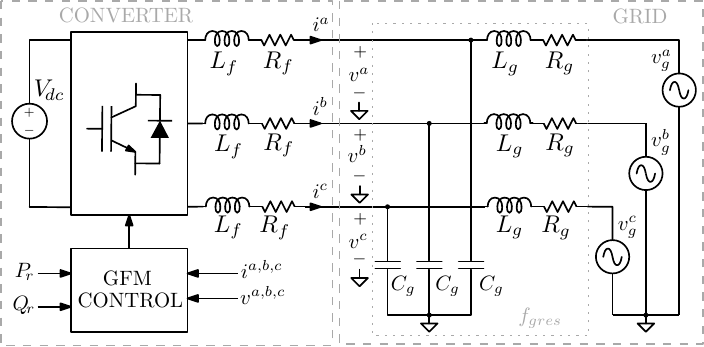}
	\caption{A grid-connected scenario used for time-domain stability tests.}
	\label{fig:scenario}
\end{figure}

\begin{figure}[!t]
	\centering
	\includegraphics[width=\linewidth,trim={0 0 0.5cm 0},clip]{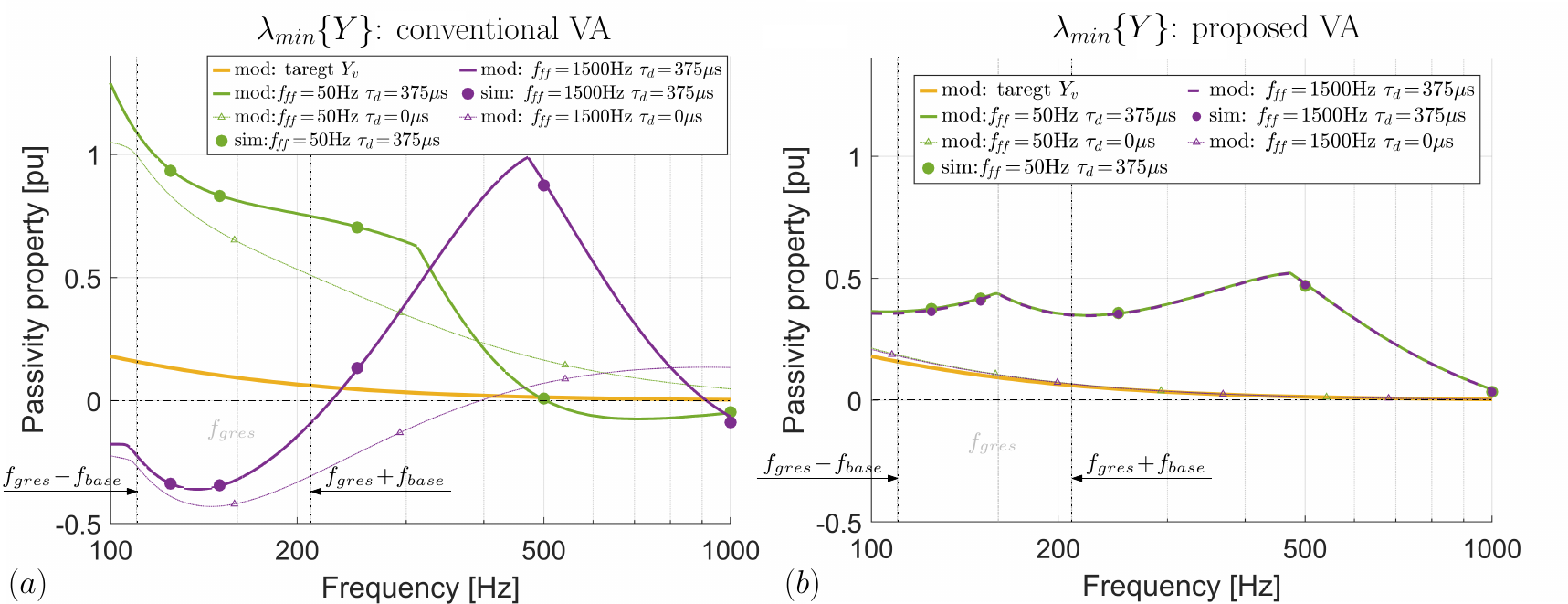}
	\caption{High-frequency passivity properties od the admittances from (a): Fig. \ref{fig:y_ap1} with conventional VA; (b) Fig. \ref{fig:y_ap2} with proposed VA approach.}
	\label{fig:lamda_ap12}
\end{figure}

\begin{figure}[!t]
	\centering
	\includegraphics[width=\linewidth,trim={0 0 0.4cm 0},clip]{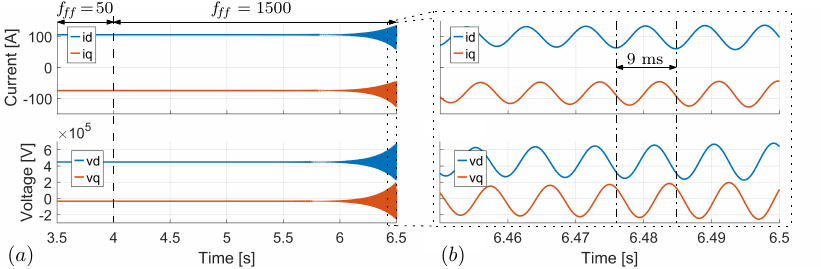}
	\caption{$Dq$ frame currents and voltages of the converter from Table I ($\tau_d = 375$ $\mu$s) and Fig. \ref{fig:bd} with the conventional VA control in the scenario from Fig. \ref{fig:scenario} $f_\mathrm{gres} = 160$ Hz (a) Response to the step change (at $t = 4$ s) from $f_\mathrm{ff} = 50$ Hz to $f_\mathrm{ff} = 1500$ Hz. (b) zoomed-in waveforms of (a).}
	\label{fig:destab_ap1}
\end{figure}

To illustrate this, a scenario from Fig. \ref{fig:scenario} is realized in simulation. The grid features an anti-resonance formed by $L_g$ and $C_g$. The value of $L_g$ that corresponds to the grid's short-circuit ratio (SCR) of $18$ is used, while $C_g$ that corresponds to $1.75$ pu is used\footnote{The selected value of the capacitance is inspired by the parameters in the HORSEA2 offshore wind-farm connection scenario, where capacitance of $1.75$ pu, corresponding to $60$ km submarine cable length, was causing destabilizing interactions between the GFM STATCOM and the grid.}. Given the base values from Table I, $L_g = 282$ mH and $C_g = 3.5$ $\mu$F, which yields the grid's anti-resonant frequency of $f_\mathrm{gres} = \frac{1}{2\pi\sqrt{L_gC_g}} \approx 160$ Hz, corresponding to $110$ Hz and $210$ Hz in the $dq$ frame. The anti-resonance is damped by the grid inductor's series resistance $R_g = 4.44$ $\Omega$, which corresponds to $0.05L_g$ [pu]. The converter features parameters from Table I and control from Fig. \ref{fig:bd}(a) with conventional VA control. Its performance is evaluated for $f_\mathrm{ff} = 50$ Hz and $f_\mathrm{ff} = 1500$ Hz. The converter's admittance passivity properties in these two cases are compared in Fig. \ref{fig:lamda_ap12}(a), where frequency responses from analytical model and simulation are shown. It is clearly seen that at $f_\mathrm{gres} \pm 50$ Hz, which is marked by dash-dotted vertical lines in Fig. \ref{fig:lamda_ap12}(a), $\lambda_\mathrm{min} < 0$ when $f_\mathrm{ff} = 1500$ Hz, while $\lambda_\mathrm{min} > 0$ with $f_\mathrm{ff} = 50$ Hz. Consequently, in the scenario from Fig. \ref{fig:scenario}, provided that the grid's passive damping is sufficiently low \cite{harneforsInputAdmittanceCalculationShaping2007,harneforsPassivityBasedControllerDesign2015}, an instability is expected when $f_\mathrm{ff} = 1500$ Hz, while a stable response is expected when $f_\mathrm{ff} = 50$ Hz. The results in Fig. \ref{fig:destab_ap1} validate these passivity-based instability risk predictions. There, the $dq$-frame currents and voltages at the point of common coupling are shown in response to the step change in $f_\mathrm{ff}$ from $50$ Hz to $1500$ Hz. An instability arises when $f_\mathrm{ff} = 1500$ Hz; as can be seen from Fig. \ref{fig:destab_ap1}(b), the frequency of the oscillations is around $f_\mathrm{gres} = 110$ Hz, which corresponds to the grid's anti-resonant frequency in the $dq$ frame.

This example illustrates how, depending on the cut-off frequency of the low-pass filter in the voltage feedforward path, imprecise admittance shaping of the conventional VA control can be detrimental to system stability. Although the previously analyzed instability can be overcome by utilizing a sufficiently low value of $f_\mathrm{ff}$, this worsens admittance passivity properties in the sub-synchronous range, as shown in Fig. \ref{fig:sweep_2}(a). Moreover, outside the influence of outer-loops, the conventional VA approach fails to emulate target admittance, thereby limiting the GFM's capability to support the grid.

Thus, it is of interest to examine alternative VA methods that could ensure precise admittance emulation in a wide frequency range outside the influence of outer-loops, regardless of the voltage feedforward and the current controller designs. As elaborated in the next section, the proposed VA approach tackles the above outlined challenges and overcomes the limitations of the conventional VA approach. 
	
\section{Proposed Virtual Admittance Control}\label{sec:ap2}
The proposed approach for VA controller design aims to ensure precise admittance shaping by considering the dynamics of the current-control loop and the voltage feedforward. For this, small-signal model from Section II-A is used, which is justified given the reasoning from Section III-A. Acc. to \eqref{eq:y_yo_hiref} and Fig. \ref{fig:bd}(e), the GFM’s admittance $\mathbf{Y_\mathrm{\!inner}}$ is impacted by the two terms \cite{caldognettoImpedanceSynthesisInverter2016,caldognettoPowerElectronicsBased2017}. First, $\mathbf{Y_o}(s)$ (from \eqref{eq:yo}), which represents the contribution when the current reference perturbation is nulled and corresponds to the GFM's admittance without VA control ($\mathbf{G_\mathrm{va} = \mathbf{0}}$ in Fig. \ref{fig:bd}(b)). Second, the term $\mathbf{H_{iref}}(s)$ (from \eqref{eq:hiref}) which is the closed-loop current reference tracking  transfer function matrix. Both of these two terms, $\mathbf{Y_o}(s)$ and $\mathbf{H_{iref}}(s)$, have to be compensated for when designing VA controller $\mathbf{G_\mathrm{va}}(s)$. This goal is met by imposing that $\mathbf{Y_\mathrm{\!inner}}(s)$ (from \eqref{eq:y_yo_hiref}) is equal to $\mathbf{Y_v}(s)$ (from \eqref{eq:yv}), which is in line with the internal model control \cite{garciaInternalModelControl1982}. This yields
\begin{equation}
    \mathbf{G_\mathrm{va}}(s) = \mathbf{H_{iref}}^{-1}(s) \left(\mathbf{Y_v}(s) - \mathbf{Y_o}(s)\right)
    \label{eq:Gva_prop_hiref_yo}
\end{equation}
By substituting $\mathbf{H_{iref}}$ from \eqref{eq:hiref} and $\mathbf{Y_o}$ from \eqref{eq:yo} in \eqref{eq:Gva_prop_hiref_yo}, and disregarding the delays ($\tau_d = 0$, i.e., $\mathbf{G_\mathrm{\!del}}(s) = \mathbf{I}$) to make the practical realization of $\mathbf{G_\mathrm{va}}$ feasible, the proposed VA controller's $s$-domain transfer function matrix is obtained
\begin{equation}
	\begin{aligned}
		\mathbf{G_\mathrm{va}}(s) &= \mathbf{G_\mathrm{ireg}}(s)^{-1}( \mathbf{G_\mathrm{ff}(s)}-\mathbf{I} + \\ &\quad \left(\mathbf{G_{Lf}(s)}+\mathbf{G_\mathrm{ireg}}(s) +\mathbf{G_\mathrm{dec}}\right)\mathbf{Y_v}(s) ).
	\end{aligned}
	\label{eq:Gvreg_prop}
\end{equation}
Acc. to \eqref{eq:Gvreg_prop}, the proposed VA approach inherently accounts for the dynamics of the current-control loop, as well as the voltage feedforward. Based on \eqref{eq:GLf}-\eqref{eq:Gdec}, \eqref{eq:Gvreg_dq} and \eqref{eq:Gvreg_prop}
\begin{equation}
G_\mathrm{va}^{\mathrm{dd}}(s) \! = \! G_\mathrm{va}^{\mathrm{qq}}(s) \! =  \!\frac{B_4 s^4 + B_3 s^3 + B_2 s^2 + B_1 s + B_0}{(s\!+\!\omega_{\!f\!f})(A_3 s^3 \!+ \!A_2 s^2  \!+ \!A_1 s\!+\!A_0)}
\label{eq:Gva_prop_dd_qq}
\end{equation}
\begin{equation}
G_{\!va}^{\mathrm{dq}}(s) \! = \! -G_{\!va}^{\mathrm{qd}}(s) \! = \! \frac{C_2 s^2 + C_1 s + C_0}{A_3 s^3\!+ \!A_2 s^2\!+ \!A_1 s\!+ \!A_0}
\label{eq:Gva_prop_dq_qd}
\end{equation}
where 
\begin{gather*}
A_3 = k_p L_v^2 \\
A_2 = k_i L_v^2 + 2k_p R_v L_v \\
A_1 = k_p R_v^2 + k_p L_v^2 \omega_g^2 + 2k_i L_v R_v \\
A_0 = k_i R_v^2 + k_i L_v^2 \omega_g^2 \\
B_4 = L_f L_v - L_v^2 \\
B_3 = L_f(L_v \omega_\mathrm{ff} + R_v) + (R_f + k_p)L_v - 2L_v R_v \\
B_2 \! = \! L_f R_v \omega_\mathrm{ff} \!+ \! (R_f \!+ \! k_p)(L_v\omega_{\!f\!f} \!+ \! R_v) \!+ \! k_i L_v \!-\! R_v^2 \!- \! L_v^2 \omega_g^2 \\
B_1 = (R_f + k_p)R_v \omega_\mathrm{ff} + k_i(L_v \omega_\mathrm{ff} + R_v) \\
B_0 = k_i R_v \omega_\mathrm{ff}
\end{gather*}
\begin{gather*}
C_2 = L_f L_v \omega_g \\
C_1 = k_p L_v \omega_g + L_v R_f \omega_g \\
C_0 = k_i L_v \omega_g.
\end{gather*}
Therefore,  as given by \eqref{eq:Gva_prop_dd_qq} and \eqref{eq:Gva_prop_dq_qd}, $s$-domain implementation of the proposed VA controller in the $dq$ frame involves two fourth-order and two third-order transfer functions. As for digital, implementation any of the standard discretization methods can be used. In this article, Tustin method is used.

\begin{figure*}[!t]
	\centering
	\includegraphics[width=\linewidth,trim={0 0 0.5cm 0},clip]{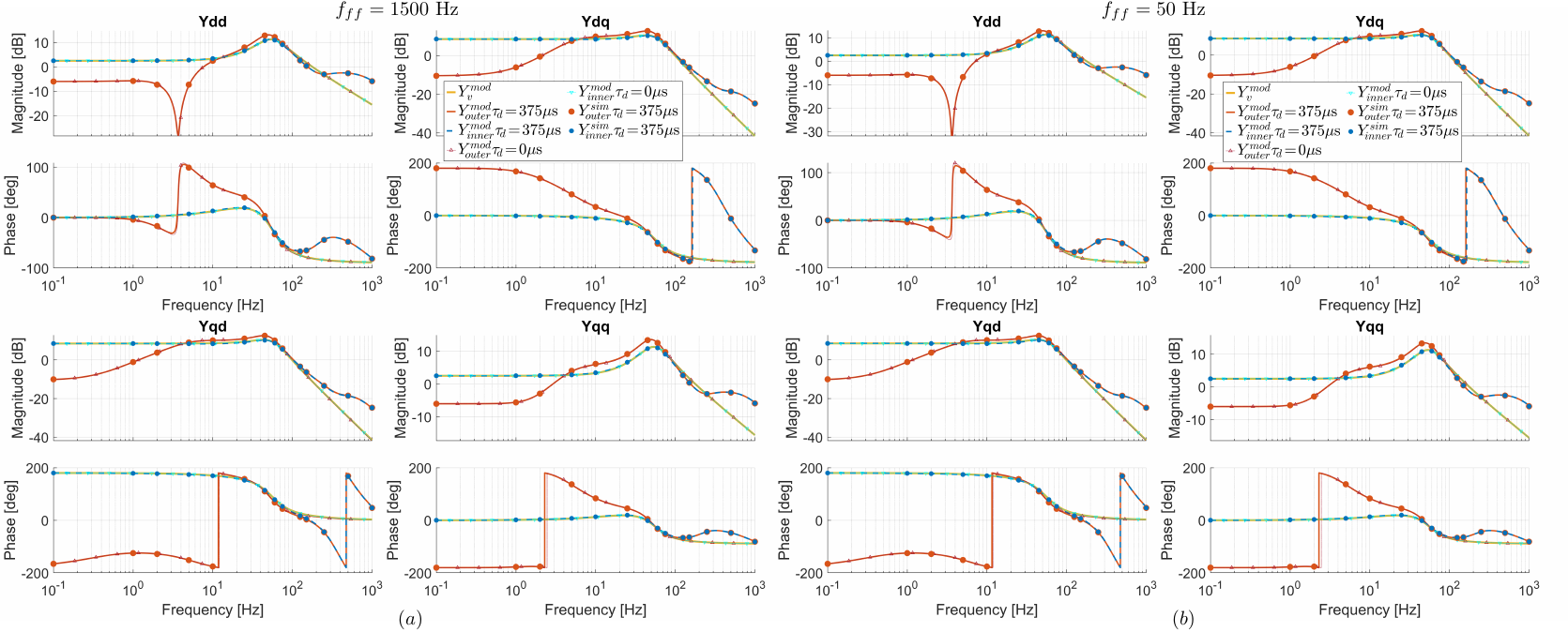}
	\caption{Normalized admittance frequency responses of the converter from Fig. \ref{fig:bd} and Table I when proposed VA approach is used: (a) $f_\mathrm{ff} = 1500$ Hz; (b) $f_\mathrm{ff} = 50$ Hz. Comparison between results from simulation (dot markers) and analytical model (lines) wit/out outer loops ($P_{r} = 0.5$ pu and $Q_{r} = 0.3$ pu).}
	\label{fig:y_ap2}
\end{figure*}

\subsection{Frequency Domain Accuracy Characterization}
To evaluate the performance of this proposed VA approach, GFM from Fig. \ref{fig:bd} and Table I is considered. Fig. \ref{fig:y_ap2}(a) and (b) show the GFM's admittance when $f_\mathrm{ff} = 1500$ Hz and $f_\mathrm{ff} = 50$ Hz, respectively. The results from  MATLAB/Simulink switching-period-averaged simulation with discrete-time control implementation and those from the analytical model with \eqref{eq:y_outer} and without outer loops \eqref{eq:y} are compered for $\tau_d = 0$ and $\tau_d = 375$ $\mu$s. Frequency responses of the target admittance $\mathbf{Y_v}$ (from \eqref{eq:yv}) are also shown. 
Without any delays ($\tau_d = 0$), outside the influence of outer loops, the proposed VA control yields perfect match between the GFM's admittance and the target admittance. Even with non-negligible delays, compared to the conventional VA approach, the proposed VA approach significantly widens the range in which the GFM's admittance is equal to the target admittance. This can be clearly seen by comparing the results from Fig. \ref{fig:y_ap2} and Fig. \ref{fig:y_ap1}. Moreover, contrary to the conventional VA approach, outside the influence of outer-loops, performance of the proposed VA approach remains the same regardless of the voltage feedforward cut-off frequency $f_\mathrm{ff}$ and the current controller parameters. This is because the proposed VA controller is designed to inherently account for, i.e. counteract these effects, by relying on internal model control.

To further illustrate the impact of the voltage feedforward with the proposed VA approach, Fig. \ref{fig:sweep}(b) is provided. There, the converter's admittance frequency responses obtained from the analytical model with outer loops \eqref{eq:y_outer} are plotted for $25$ different values of $f_\mathrm{ff}$. Compared to the results with the conventional VA approach from Fig. \ref{fig:sweep}(a), with the proposed VA approach, the frequency range where $f_\mathrm{ff}$ affects admittance frequency responses is significantly reduced. 

\subsection{Admittance Passivity and System Stability}

To show how wide-band admittance shaping achieved by the proposed VA control is advantageous for GFM passivity, $\lambda_\mathrm{min}$ of the admittances from Fig. \ref{fig:sweep}(b) is plotted in Fig. \ref{fig:sweep_2}(b). Compared to Fig. \ref{fig:sweep_2}(a), not only is the low-frequency negative dip of $\lambda_\mathrm{min}$ reduced, but also is the high-frequency active region removed. 

To further illustrate how the proposed VA approach is advantageous for passivity in the harmonic range, Fig. \ref{fig:lamda_ap12}(b) comapres $\lambda_\mathrm{min}$ in the range $100$-$1000$ Hz for $f_\mathrm{ff} = 50$ Hz and $f_\mathrm{ff} = 1500$ Hz. Contrary to Fig. \ref{fig:lamda_ap12}(a), that corresponds to the conventional VA approach, $\lambda_\mathrm{min}$ with the proposed VA approach does not become negative around $100$-$200$ Hz neither when $f_\mathrm{ff} = 50$ Hz nor when $f_\mathrm{ff} = 1500$ Hz. Moreover, this holds for an arbitrary value of $f_\mathrm{ff}$, as its impact is "inverted" in the proposed VA controller design. Consequently, the GFM's robustness to grid disturbances in the harmonic range is significantly enhanced.

\begin{figure}[!t]
	\centering
	\includegraphics[width=\linewidth,trim={0 0 0.5cm 0},clip]{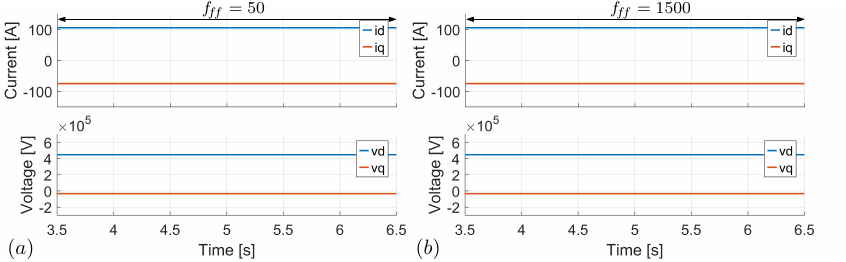}
	\caption{$Dq$ frame currents and voltages of the converter from Table I ($\tau_d =375$ $\mu$s) and Fig. \ref{fig:bd} with the proposed VA control in scenario from Fig. \ref{fig:scenario} $f_\mathrm{gres} = 160$: (a) $f_\mathrm{ff} = 50$ Hz; (b) $f_\mathrm{ff} = 1500$ Hz.}
	\label{fig:stab_ap2}
\end{figure}

To demonstrate this, a scenario from Fig. \ref{fig:scenario} is tested in the Simulink simulation model. Same as before, the grid features an anti-resonance at $f_\mathrm{gres} \approx 160$ Hz. The converter employs the proposed VA control and its performance is evaluated for $f_\mathrm{ff} = 1500$ Hz and $f_\mathrm{ff} = 50$ Hz. Note that, as given by \eqref{eq:Gva_prop_dd_qq}-\eqref{eq:Gva_prop_dq_qd}, with the proposed approach, the VA controller parameters depend on the value of $f_\mathrm{ff}$, which in the considered control architecture cannot be tuned online. Thus, a single grid-connected stability test under the change of $f_\mathrm{ff}$ is not suitable, as the VA controller parameters would not have been updated after changing $f_\mathrm{ff}$. Therefore, two set of measurements are performed, one with $f_\mathrm{ff} = 1500$ Hz and the other one with $f_\mathrm{ff} = 50$ Hz. The results are shown in Fig. \ref{fig:stab_ap2}(a) and Fig. \ref{fig:stab_ap2}(b), respectively, where the $dq$ frame current and voltages at the GFM's point of connection are plotted. As can be seen, the stable operation is retained regardless of the value of $f_\mathrm{ff}$. This is aligned with the stability implications deduced from the passivity properties on Fig. \ref{fig:lamda_ap12}(b). This demonstrates how, by ensuring wide-band admittance shaping, the proposed VA control outperforms the state-of-the-art VA control in terms of GFM's impact on system stability and robustness.

\section{Experimental Validation}\label{sec:exp}

To validate the above presented methodology experimentally, the setup from Fig. \ref{fig:setup} is built. It features two three phase voltage source converters (denoted by (2) and (3) in Fig. \ref{fig:setup}), which are realized by Imperix all-in-one programmable converters TPI8032. Both converters have the control system from Fig. \ref{fig:bd} and the parameters from Table II. 

Two sets of measurements are performed, which are explained in the following two subsections. For both sets of measurements, the converter (2) is considered as a GFM whose properties are to be evaluated. As such, it was tested for both the conventional and for the proposed VA control. On the other side, the converter (3) is used to emulate the grid and for perturbation injection. The converter (3) always employs the proposed VA control, given its assets discussed in Section IV.

\subsection{Frequency Domain Accuracy Characterization}
\begin{figure}[!t]
	\centering
	\includegraphics[width=\linewidth]{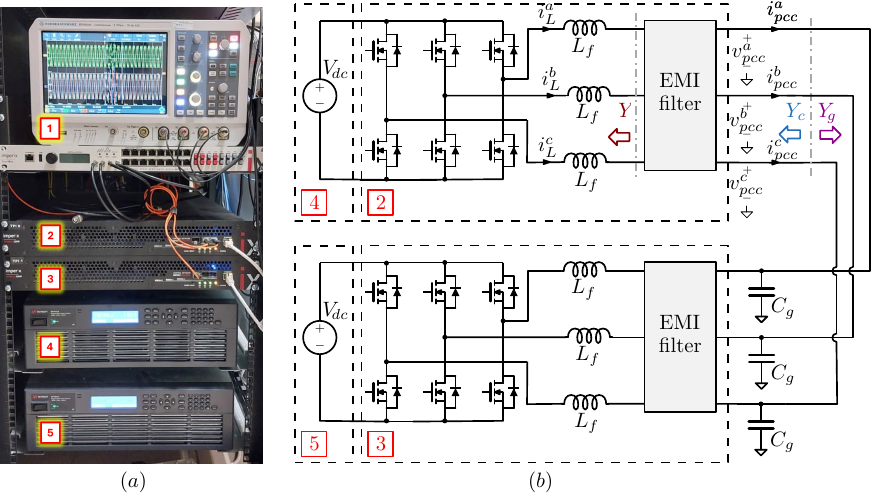}
	\caption{(a) Picture and (b) block diagram of the setup used for experimental validation: (1) oscilloscope; (2)-(3) TPI8032 converter; (4)-(5) dc supplies.}
	\label{fig:setup}
\end{figure}

The goal of the first set of measurements was to characterize accuracy of admittance emulation in frequency domain. For this, the capacitance $C_g$ from Fig. \ref{fig:setup} is not connected at the point-of-common-coupling (PCC) and, referring to Table II, $S_\mathrm{base} = 1$ kW and $f_\mathrm{ff} = \{ 0, \infty \}$ is used. Note that, $f_\mathrm{ff} = 0$ corresponds to not using the voltage feedforward, while $f_\mathrm{ff} = \infty$ corresponds to using the voltage feedforward without any low-pass filter. Frequency responses of the converter's admittance (denoted by $\mathbf{Y_{c}}$ in Fig. \ref{fig:setup}) are experimentally measured in the following way. A sinusoidal perturbation is injected via converter (3) from Fig. \ref{fig:setup}, where, for each perturbation frequency of interest, two independent perturbation voltages are injected, one along $d$ axis and the other along $q$ axis. After each perturbation injection, once the steady-state has been reached, the variables of interest, i.e., the converter's phase currents ($i_{pcc}^a$, $i_{pcc}^b$, $i_{pcc}^c$ in Fig. \ref{fig:setup}) and voltages ($v_{pcc}^a$, $v_{pcc}^b$, $v_{pcc}^c$ in Fig. \ref{fig:setup}) are acquired at the rate equal to the converter's control algorithm execution rate $f_s$. Then, they are transformed to the $dq$ frame and a Fast Fourier Transform is performed in real time, to determine current and voltage components at the perturbation frequency. By dividing them, frequency responses of the converter's admittance $\mathbf{Y_{c}}$ from Fig. \ref{fig:setup} are obtained. These are then imported in MATLAB where the a priori measured EMI filter's admittance frequency responses $\mathbf{Y_{emi}}$ are "subtracted", so that the frequency responses of the converter's admittance before this filter are obtained, which is $\mathbf{Y} = \mathbf{Y_c} - \mathbf{Y_{emi}}$. 

\begin{table}[t!]
	\caption*{Table II: Parameters used for experimental validation.}
	\centering
	\begin{tabular}{lccc}
		\midrule\midrule
		Parameter & label & value & unit\\
		\midrule
		Base power & $S_\mathrm{base}$ & $\{1,\color{gray}{5} \color{black}\}$ &  kW  \\     
		Base voltage (ph-ph) & $V_\mathrm{base}$ & $100$ &  Vrms  \\
		Base (grid) frequency & $f_\mathrm{base}$ & $50$ &  Hz  \\
		DC link voltage & $V_{dc}$ & $200$ &  V  \\
		Filter inductance & $L_f$ & $950$ & $\mu$H \\
		Filter resistance & $R_f$ & $115$ & m$\Omega$. \\
		Switching frequency & $f_{pwm}$ & $25$ & kHz \\
		Sampling frequency & $f_{s}$ & $50$ & kHz \\
		Virtual inductance & $L_v^{pu}$ & $0.3$ & p.u. \\
		Virtual resistance & $R_v^{pu}$ & $0.03$ & p.u. \\
		Current loop bandwidth & $f_\mathrm{ireg}$ & $2.5$ & kHz \\
		Feedforward cut-off frequency & $f_{\!f\!f}$ & $\{0, 150, 1500, \infty\}$ & Hz \\
		Control loop delay & $\tau_{d}$ & $ 30$ & $\mu$s \\
        Outer (PQ) loop bandwidth & $f_{\mathrm{pqreg}}$ & $1$ & Hz \\
		\midrule\midrule
		
	\end{tabular}
\end{table}

\begin{figure}[!t]
	\centering
	\includegraphics[width=\linewidth,trim={0 0 0.65cm 0},clip]{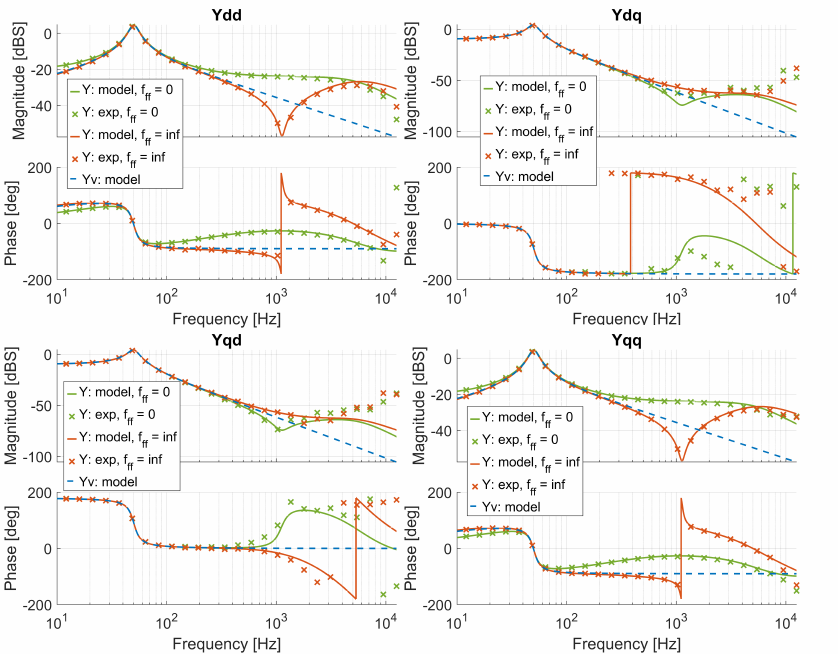}
	\caption{Experimental admittance frequency responses of the GFM from Table II ($S_{base} = 1$ kW) and Fig. \ref{fig:bd} with conventional VA approach.}
	\label{fig:exp_ap1_y}
\end{figure}

\begin{figure}[!t]
	\centering
	\includegraphics[width=\linewidth,trim={0 0 0.65cm 0},clip]{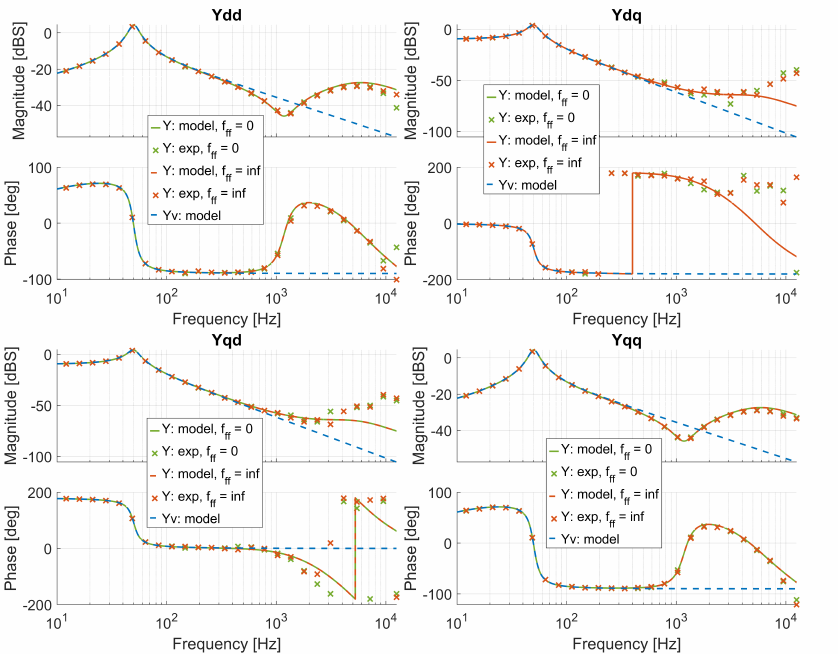}
	\caption{Experimental admittance frequency responses of the GFM from Table II ($S_{base} = 1$ kW) and Fig. \ref{fig:bd} with proposed VA approach.}
	\label{fig:exp_ap2_y}
\end{figure}

The resulting admittance frequency responses are shown in Fig. \ref{fig:exp_ap1_y} and Fig. \ref{fig:exp_ap2_y}, which correspond to the case when the converter employs the conventional and the proposed VA control, respectively. As can be seen, the experimentally obtained data is in excellent agreement with the data obtained using analytical model. In addition, the presented results demonstrate how the proposed VA control is advantageous for achieving target admittance frequency response. Particularly, with respect to the conventional, the proposed VA control extends the frequency range where the converter's admittance matches the target. Moreover, performance of the proposed VA control is not dependent on the value of $f_\mathrm{ff}$, which as addressed below has positive impact on stability.

\subsection{Impact of Admittance Shaping Accuracy on Stability}

The goal of the second set of measurements was to experimentally demonstrate the impact of VA control on system stability. For this set of measurements, referring to Table II, $S_\mathrm{base} = 5$ kW and $f_\mathrm{ff} = \{ 150, 1500\}$ Hz is used. The same as before, the converter denoted by (2) is used as a GFM under test. The converter (3) is used to emulate the grid with inductive impedance, featuring inductance $L_g = 1.9$ mH and resistance $R_g = 0.12$ $\Omega$. The capacitance $C_g = 170$ $\mu$F is connected at the PCC to form an anti-resonance (at around $f_\mathrm{gres} = 280$ Hz) that may impair system stability. In this way, a scenario similar to the scenario from Fig. \ref{fig:scenario} is created, which serves as a simple and illustrative test-case for evaluating robustness of the GFM.

\begin{figure}[!t]
	\centering
	\includegraphics[width=\linewidth,trim={0 0 0.5cm 0},clip]{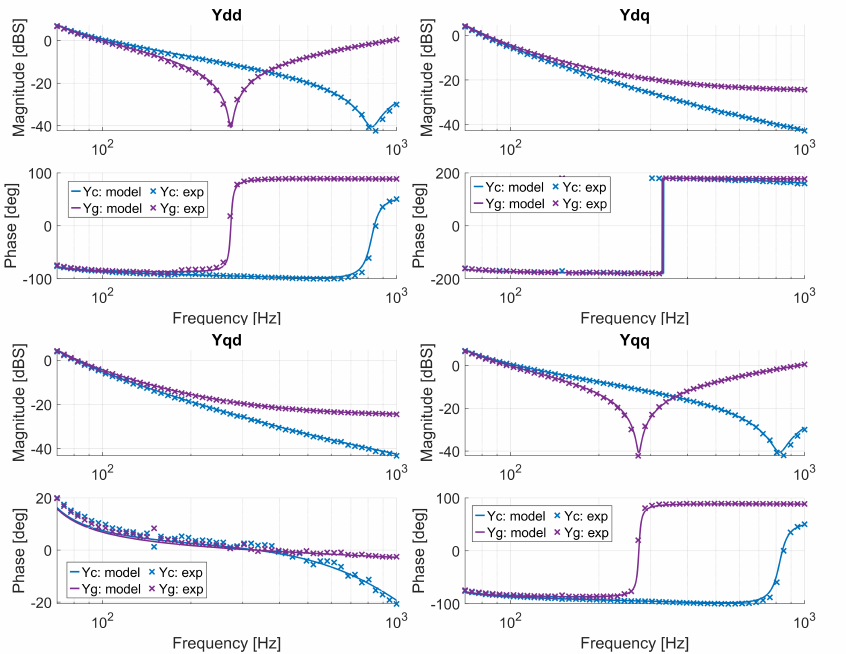}
	\caption{Experimental frequency responses of the equivalent grid admittance ($Y_g$ from Fig. \ref{fig:setup}) and the GFM admittance ($Y_c$ from Fig. \ref{fig:setup}) in case the conventional VA control with $f_\mathrm{ff} = 1500$ Hz is used for the converter under test (marked by (2) on Fig. \ref{fig:setup}).}
	\label{fig:exp_yyg}
\end{figure}

First, it was of interest to show how depending on the value of $f_\mathrm{ff}$ an instability may arise if the conventional VA control is used. Thus, the case where converter (2) from Fig. \ref{fig:setup} employs conventional VA control is first tested. The frequency responses of the converter's and the grid's admittances, marked by $\mathbf{Y_c}$ and $\mathbf{Y_g}$ respectively in Fig. \ref{fig:setup}(b), are shown in Fig. \ref{fig:exp_yyg} for the case when $f_\mathrm{ff} = 1500$ Hz, which is critical from the stability point-of-view. Since, as can be seen from Fig. \ref{fig:exp_yyg}, the experimentally measured results match excellently those obtained from the analytical model, a subsequent impedance-based stability analysis is performed using the analytical model.

\begin{figure}[!t]
	\centering
	\includegraphics[width=\linewidth]{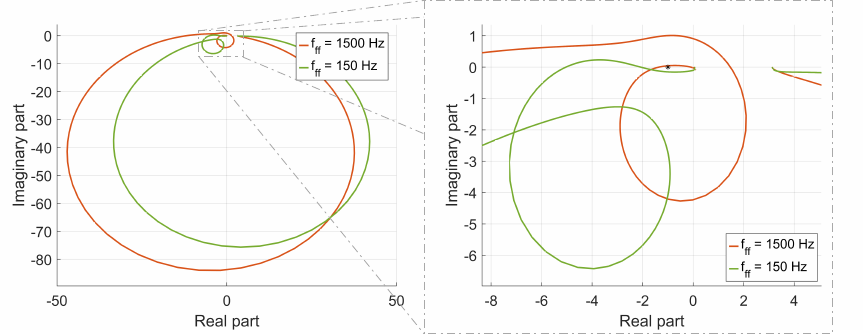}
	\caption{Nyquist plots of $L_m$ from \eqref{eq:Lm} for the system from Fig. \ref{fig:setup} and Table II ($S_{base} = 5$ kW) when conventional VA approach is used for converter (2).}
	\label{fig:exp_ap1_nyq}
\end{figure}

\begin{figure*}[!t]
	\centering
	\includegraphics[width=\linewidth,trim={0 0 0 0.2cm},clip]{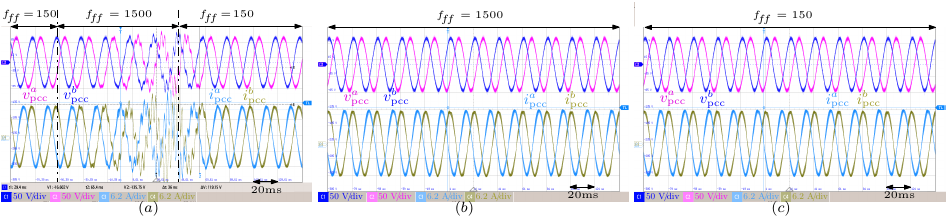}
	\caption{Experimentally measured PCC phase current and voltages of the circuit from Fig. \ref{fig:setup} in case the converter (2) uses: (a) conventional VA approach with the step change from $f_\mathrm{ff} = 150$ Hz to $f_\mathrm{ff} = 1500$ Hz and vice versa; proposed VA approach with (b) $f_\mathrm{ff} = 1500$ Hz and (c) $f_\mathrm{ff} = 150$ Hz.}
	\label{fig:exp_scope}
\end{figure*}

For this, the minor loop gain is defined as $\mathbf{Y_g}^{-1}\mathbf{Y_c}$ and a determinant-based variant of the Generalized Nyquist criterion is applied. This involves applying the Nyquist criterion to
\begin{equation}
	L_m = \det(\mathbf{I} + \mathbf{Y_g}^{-1}\mathbf{Y_c}) -1
	\label{eq:Lm}
\end{equation}
where $\det$ refers to the determinant. The resulting Nyquist plots of $L_m$ obtained when the conventional VA control is used with $f_\mathrm{ff} = 1500$ Hz and $f_\mathrm{ff} = 150$ Hz are shown in Fig. \ref{fig:exp_ap1_nyq}. Given that the minor loop gain does not feature right half-plane poles for neither $f_\mathrm{ff} = 1500$ Hz nor $f_\mathrm{ff} = 150$ Hz, as validated by the analytical model, closed-loop stability can simply be assessed by checking whether the plotted $L_m$ curves encircle the critical (-1,0) point. Consequently, as can be seen from Fig. \ref{fig:exp_ap1_nyq}, a stable response is predicted for $f_\mathrm{ff} = 150$ Hz, while an instability is predicted for $f_\mathrm{ff} = 150$ Hz. To validate this, time domain stability test is performed where the step change of the feedforward cut-off frequency $f_\mathrm{ff}$ from $f_\mathrm{ff} = 150$ Hz to $f_\mathrm{ff} = 1500$ Hz and vise versa is imposed. The current and voltage waveforms in response to this transient are shown in Fig. \ref{fig:exp_scope}(a). As predicted, when $f_\mathrm{ff} = 1500$ Hz an instability arises. The results clearly demonstrate how conventional VA control can be detrimental for system stability.

\begin{figure}[!t]
	\centering
	\includegraphics[width=\linewidth]{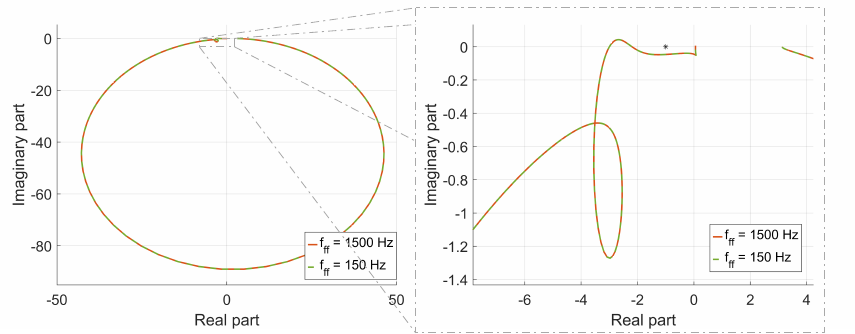}
	\caption{Nyquist plots of $L_m$ from \eqref{eq:Lm} for the system from Fig. \ref{fig:setup} and Table II ($S_{base} = 5$ kW) when proposed VA approach is used for converter (2).}
	\label{fig:exp_ap2_nyq}
\end{figure}

To illustrate how the proposed VA control overcomes the limitation of the conventional one, the scenario from Fig. \ref{fig:setup} is now considered with the converter (2) employing the proposed VA approach. The resulting Nyquist plots of $L_m$ from \eqref{eq:Lm} obtained when the proposed VA approach is used with $f_\mathrm{ff} = 1500$ Hz and $f_\mathrm{ff} = 150$ Hz are shown in Fig. \ref{fig:exp_ap2_nyq}. As can be seen, stable operation is predicted regardless of the value of $f_\mathrm{ff}$. The corresponding experimentally measured time domain stability test results are shown in Fig. \ref{fig:exp_scope}(b)-(c). As can be seen, regardless of the value of $f_\mathrm{ff}$, stable operation is ensured. This attests to the ability of the proposed VA control to ensure robust GFM performance, which is due to the wide-band admittance shaping that this approach achieves.

\section{Conclusions}\label{sec:co}

By examining the capability to achieve a target admittance frequency response, this work investigates the impact of admittance shaping on the stability of grid-forming converters, considering two virtual admittance approaches: a state-of-the-art one and a novel internal-model-control-based approach that accounts for the effect of the current control loop and the voltage feedforward. The results demonstrate that, in contrast to the conventional approach, the proposed approach enables the grid-forming converter to accurately reproduce the target immittance over a wide frequency range, outside the influence of outer loops. Furthermore, it has been shown that when a low-pass filter with a high cut-off frequency is employed in the voltage feedforward path, the state-of-the-art VA approach may lead to instability in the harmonic range. On the other hand, the proposed approach guaranties stable operation irrespective of the selected feedforward's cut-off frequency. The effectiveness and practical feasibility of the proposed methodology have been confirmed through experimental validation on a laboratory-scale prototype featuring three-phase voltage-source converters.


\bibliographystyle{IEEEtran}
\bibliography{bib/IEEEabrv,bib/mybibliography.bib}

@STRING{IEEE_J_PWRE       = "{IEEE} Trans. Power Electron."}

@STRING{IEEE_J_IE         = "{IEEE} Trans. Ind. Electron."}

@STRING{IEEE_J_IA         = "{IEEE} Trans. Ind. Appl."}

@article{garciaInternalModelControl1982,
  title = {Internal Model Control. {{A}} Unifying Review and Some New Results},
  author = {Garcia, Carlos E. and Morari, Manfred},
  year = {1982},
  month = apr,
  journal = {Industrial \& Engineering Chemistry Process Design and Development},
  volume = {21},
  number = {2},
  pages = {308--323},
  publisher = {American Chemical Society},
  issn = {0196-4305},
  doi = {10.1021/i200017a016},
  urldate = {2026-07-19}
}

@article{guoImpedanceAnalysisStabilization2021,
  title = {Impedance {{Analysis}} and {{Stabilization}} of {{Virtual Synchronous Generators With Different DC-Link Voltage Controllers Under Weak Grid}}},
  author = {Guo, Jian and Chen, Yandong and Wang, Lei and Wu, Wenhua and Wang, Xiangyu and Shuai, Zhikang and Guerrero, Josep M.},
  year = {2021},
  month = oct,
  journal = IEEE_J_PWRE,
  volume = {36},
  number = {10},
  pages = {11397--11408},
  issn = {1941-0107},
  doi = {10.1109/TPEL.2021.3070038},
  urldate = {2026-07-18}
}

@article{imgartExternalInertiaEmulation2024,
	title = {External {{Inertia Emulation}} to {{Facilitate Active-Power Limitation}} in {{Grid-Forming Converters}}},
	author = {Imgart, Paul and Narula, Anant and Bongiorno, Massimo and Beza, Mebtu and Svensson, Jan R.},
	year = {2024},
	month = nov,
	journal = IEEE_J_IA,
	volume = {60},
	number = {6},
	pages = {9145--9156},
	issn = {1939-9367},
	doi = {10.1109/TIA.2024.3443792},
	urldate = {2025-10-04}
}

@misc{AEMO2024,
	author = {Australian Energy Market Operator},
	title = {Voluntary Specification for Grid-forming Inverters: Core Requirements Test Framework},
	year = {2024},
	month = {Jan},
	url = {https://www.aemo.com.au/-/media/files/initiatives/engineering-framework/2023/grid-forming-inverters-jan-2024.pdf},
}

@misc{GBGF2023,
	author = {{National Grid ESO}},
	title = {{GBGF} Best Practice Guide – April 2023},
	year = {2023},
	url = {https://www.scribd.com/document/681971716/GBGF-Best-Practice-Guide-April-2023},
}

@article{matevosyanGridFormingInvertersAre2019,
	title = {Grid-{{Forming Inverters}}: {{Are They}} the {{Key}} for {{High Renewable Penetration}}?},
	shorttitle = {Grid-{{Forming Inverters}}},
	author = {Matevosyan, Julia and Badrzadeh, Babak and Prevost, Thibault and Quitmann, Eckard and Ramasubramanian, Deepak and Urdal, Helge and Achilles, Sebastian and MacDowell, Jason and Huang, Shun Hsien and Vital, Vijay and O'Sullivan, Jon and Quint, Ryan},
	year = {2019},
	month = nov,
	journal = {IEEE Power and Energy Magazine},
	volume = {17},
	number = {6},
	pages = {89--98},
	issn = {1558-4216},
	doi = {10.1109/MPE.2019.2933072},
	urldate = {2025-10-04}
}

@inproceedings{caldognettoImpedanceSynthesisInverter2016,
	title = {Impedance {S}ynthesis by {I}nverter {C}ontrol for {A}ctive {L}oads in {A}nti-{I}slanding {T}estbenches},
	booktitle = {2016 {{IEEE Energy Conversion Congress}} and {{Exposition}} ({{ECCE}})},
	author = {Caldognetto, Tommaso and Dalla Santa, Luca and Magnone, Paolo and Mattavelli, Paolo},
	year = {2016},
	month = sep,
	pages = {1--7},
	doi = {10.1109/ECCE.2016.7855184},
	urldate = {2025-10-04}
}

@article{caldognettoPowerElectronicsBased2017,
	title = {Power {{Electronics Based Active Load}} for {{Unintentional Islanding Testbenches}}},
	author = {Caldognetto, Tommaso and Santa, Luca Dalla and Magnone, Paolo and Mattavelli, Paolo},
	year = {2017},
	month = jul,
	journal = IEEE_J_IA,
	volume = {53},
	number = {4},
	pages = {3831--3839},
	issn = {1939-9367},
	doi = {10.1109/TIA.2017.2694384},
	urldate = {2025-10-04}
}

@misc{entso-eHighPenetrationPower2020,
	title = {High {{Penetration}} of {{Power Electronic Interfaced Power Sources}} and the {{Potential Contribution}} of {{Grid Forming Converters}}},
	author = {{ENTSO-E}},
	year = {2020}
}

@article{zhaoExploringDampingEffect2025,
  title = {Exploring {{Damping Effect}} of {{Inner Control Loops}} for {{Grid-Forming VSCs}}},
  author = {Zhao, Liang and Wang, Xiongfei and Jin, Zheming},
  year = {2025},
  journal = {IEEE Open Journal of Power Electronics},
  volume = {6},
  pages = {1595--1608},
  issn = {2644-1314},
  doi = {10.1109/OJPEL.2025.3614708},
  urldate = {2026-06-04}
}

@article{vattakkuniComparativeAssessmentTypical2023,
  title = {Comparative Assessment of Typical Control Realizations of Grid Forming Converters Based on Their Voltage Source Behaviour},
  author = {Vatta Kkuni, Kanakesh and Mohan, Sibin and Yang, Guangya and Xu, Wilsun},
  year = {2023},
  month = dec,
  journal = {Energy Reports},
  volume = {9},
  pages = {6042--6062},
  issn = {2352-4847},
  doi = {10.1016/j.egyr.2023.05.073},
  urldate = {2026-06-04}
}

@inproceedings{rodriguezControlGridconnectedPower2013,
	title = {Control of Grid-Connected Power Converters Based on a Virtual Admittance Control Loop},
	booktitle = {2013 15th {{European Conference}} on {{Power Electronics}} and {{Applications}} ({{EPE}})},
	author = {Rodriguez, Pedro and Candela, Ignacio and Citro, Costantino and Rocabert, Joan and Luna, Alvaro},
	year = {2013},
	month = sep,
	pages = {1--10},
	doi = {10.1109/EPE.2013.6634621},
	urldate = {2025-10-04}
}

@article{cardozoPromisesChallengesGrid2024,
  title = {Promises and Challenges of Grid Forming: {{Transmission}} System Operator, Manufacturer and Academic View Points},
  shorttitle = {Promises and Challenges of Grid Forming},
  author = {Cardozo, Carmen and Prevost, Thibault and Huang, Shun-Hsien and Lu, Jingwei and Modi, Nilesh and Hishida, Masaya and Li, Xiaoming and Abdalrahman, Adil and Samuelsson, P{\"a}r and Cutsem, Thierry Van and Laba, Yorgo and Lamrani, Yahya and Colas, Frederic and Guillaud, Xavier},
  year = {2024},
  month = oct,
  journal = {Electric Power Systems Research},
  volume = {235},
  pages = {110855},
  issn = {0378-7796},
  doi = {10.1016/j.epsr.2024.110855},
  urldate = {2026-06-04}
}

@article{avdiajPassivityBasedDesignMethodology2025,
  title = {Passivity-{{Based Design Methodology}} for {{Current-Controlled Grid-Forming MMCs}}},
  author = {Avdiaj, Eros and Lee, Dongyeong and Sakinci, {\"O}zg{\"u}r Can and Beerten, Jef},
  year = {2025},
  month = oct,
  journal = {IEEE Journal of Emerging and Selected Topics in Industrial Electronics},
  volume = {6},
  number = {4},
  pages = {1366--1377},
  issn = {2687-9743},
  doi = {10.1109/JESTIE.2025.3585455},
  urldate = {2026-06-04}
}

@article{eggersAccuracyStabilityAssessment2026,
  title = {Accuracy and {{Stability Assessment}} of {{Resistive-Inductive Virtual Impedances}} for {{Grid-Forming Converters With LCL Filters}}},
  author = {Eggers, Malte and {Kaufmann-B{\"u}hler}, Marius and Dieckerhoff, Sibylle},
  year = {2026},
  month = apr,
  journal = {IEEE Journal of Emerging and Selected Topics in Power Electronics},
  volume = {14},
  number = {2},
  pages = {2414--2424},
  issn = {2168-6785},
  doi = {10.1109/JESTPE.2026.3658144},
  urldate = {2026-06-03}
}

@inproceedings{huangImpactVirtualAdmittance2021,
	title = {Impact of {{Virtual Admittance}} on {{Small-Signal Stability}} of {{Grid-Forming Inverters}}},
	booktitle = {2021 6th {{IEEE Workshop}} on the {{Electronic Grid}} ({{eGRID}})},
	author = {Huang, Liang and Wu, Chao and Zhou, Dao and Blaabjerg, Frede},
	year = {2021},
	month = nov,
	pages = {1--8},
	doi = {10.1109/eGRID52793.2021.9662150},
	urldate = {2025-10-04}
}

@misc{ipec2026,
  author       = {Cvetanovic, Ruzica and Sbabo, Paolo and Stojanovic, Lazar and Mattavelli, Paolo and Bongiorno, Massimo},
  title        = {Precise {V}irtual {A}dmittance {S}ynthesis for {E}nhanced {G}rid-{F}orming {P}erformance},
  publisher    = {TechRxiv},
  note         = {IPEC-Nagasaki 2026 (IEEE ECCE Asia) preprint  doi: 10.36227/techrxiv.177138705.53518107/v1}
}

@article{pengImprovedReactanceControl2026,
  title = {An {{Improved Reactance Control}} for {{Stability Enhancement}} of {{Grid-Forming Converter Based}} on {{Series-Regulated Branch}}},
  author = {Peng, Minxuan and Sun, Jianjun and Peng, Neng and Li, Qionglin and Wang, Yi and Zha, Xiaoming and Tang, Yi},
  year = {2026},
  month = jun,
  journal = IEEE_J_IE,
  volume = {73},
  number = {6},
  pages = {8678--8690},
  issn = {1557-9948},
  doi = {10.1109/TIE.2026.3651344},
  urldate = {2026-07-29}
}

@article{zhongSynchronvertersInvertersThat2011,
  title = {Synchronverters: {{Inverters That Mimic Synchronous Generators}}},
  shorttitle = {Synchronverters},
  author = {Zhong, Qing-Chang and Weiss, George}, 
    year = {2011},
  month = apr,
  journal = IEEE_J_IE,
  volume = {58},
  number = {4},
  pages = {1259--1267},
  issn = {1557-9948},
  doi = {10.1109/TIE.2010.2048839},
  urldate = {2026-07-29}
}

@INPROCEEDINGS{anant_GFM_comparison,
  author={Narula, Anant and Bongiorno, Massimo and Beza, Mebtu},
  booktitle={2021 IEEE Energy Conversion Congress and Exposition (ECCE)}, 
  title={Comparison of Grid-Forming Converter Control Strategies}, 
  year={2021},
  volume={},
  number={},
  pages={361-368},
  doi={10.1109/ECCE47101.2021.9594941}}

@phdthesis{narulaPhD,
	title = {Grid-forming wind power plants},
	author = {Narula, Anant},
	year = {2023},
	school = {Chalmers University of Technology},
	url = {https://research.chalmers.se/publication/534815/file/534815_Fulltext.pdf}
}

@article{rossoGridFormingConvertersControl2021,
	title = {Grid-{{Forming Converters}}: {{Control Approaches}}, {{Grid-Synchronization}}, and {{Future Trends}}---{{A Review}}},
	shorttitle = {Grid-{{Forming Converters}}},
	author = {Rosso, Roberto and Wang, Xiongfei and Liserre, Marco and Lu, Xiaonan and Engelken, Soenke},
	year = {2021},
	journal = {IEEE Open Journal of Industry Applications},
	volume = {2},
	pages = {93--109},
	issn = {2644-1241},
	doi = {10.1109/OJIA.2021.3074028},
	urldate = {2025-10-04}
}

@article{tozakModelingControlGrid2024,
	title = {Modeling and {{Control}} of {{Grid Forming Converters}}: {{A Systematic Review}}},
	shorttitle = {Modeling and {{Control}} of {{Grid Forming Converters}}},
	author = {Tozak, Macit and Taskin, Sezai and Sengor, Ibrahim and Hayes, Barry P.},
	year = {2024},
	journal = {IEEE Access},
	volume = {12},
	pages = {107818--107843},
	issn = {2169-3536},
	doi = {10.1109/ACCESS.2024.3437236},
	urldate = {2025-10-04}
}

@article{leonGridFormingControllerBased2023,
	title = {Grid-{{Forming Controller Based}} on {{Virtual Admittance}} for {{Power Converters Working}} in {{Weak Grids}}},
	author = {Leon, Jose David Vidal and Tarraso, Andres and Candela, Jose Ignacio and Rocabert, Joan and Rodriguez, Pedro},
	year = {2023},
	month = jul,
	journal = {IEEE Journal of Emerging and Selected Topics in Industrial Electronics},
	volume = {4},
	number = {3},
	pages = {791--801},
	issn = {2687-9743},
	doi = {10.1109/JESTIE.2023.3244744},
	urldate = {2025-10-04}
}

@techreport{fingrid,
	type = {Fingrid},
	title = {Grid Code Specifications for Grid Energy Storage Systems},
	year = {2024},
	month = jun,
	doi = {https://www.fingrid.fi/en/grid/grid-connection-agreement-phases/grid-code-specifications/grid-energy-storage-systems/}
}

@article{harneforsInputAdmittanceCalculationShaping2007,
	title = {Input-{{Admittance Calculation}} and {{Shaping}} for {{Controlled Voltage-Source Converters}}},
	author = {Harnefors, Lennart and Bongiorno, Massimo and Lundberg, Stefan},
	year = {2007},
	month = dec,
	journal = IEEE_J_IE,
	volume = {54},
	number = {6},
	pages = {3323--3334},
	issn = {1557-9948},
	doi = {10.1109/TIE.2007.904022},
	urldate = {2024-01-17}
}

@article{harneforsPassivityBasedControllerDesign2015,
  title = {Passivity-{{Based Controller Design}} of {{Grid-Connected VSCs}} for {{Prevention}} of {{Electrical Resonance Instability}}},
  author = {Harnefors, Lennart and Yepes, Alejandro G. and Vidal, Ana and {Doval-Gandoy}, Jes{\'u}s},
  year = {2015},
  month = feb,
  journal = IEEE_J_IE,
  volume = {62},
  number = {2},
  pages = {702--710},
  issn = {1557-9948},
  doi = {10.1109/TIE.2014.2336632},
  urldate = {2026-07-28}
}

@article{liuUnifiedVoltageControl2024,
  title = {Unified {{Voltage Control}} for {{Grid-Forming Inverters}}},
  author = {Liu, Teng and Wang, Xiongfei},
  year = {2024},
  month = mar,
  journal = IEEE_J_IE,
  volume = {71},
  number = {3},
  pages = {2578--2589},
  issn = {1557-9948},
  doi = {10.1109/TIE.2023.3265055},
  urldate = {2026-07-28}
}

@article{zhaoAnalysisActiveDamping2026,
  title = {Analysis of {{Active Damping Methods}} for {{Power-Synchronization Dynamics Considering Impacts}} of {{Different Inner Loops}}},
  author = {Zhao, Liang and Wang, Xiongfei and Li, Zejie},
  year = {2026},
  journal = IEEE_J_IE,
  pages = {1--13},
  issn = {1557-9948},
  doi = {10.1109/TIE.2026.3679775},
  urldate = {2026-07-28}
}

@article{wangHarmonicStabilityPower2019,
	title = {Harmonic {{Stability}} in {{Power Electronic-Based Power Systems}}: {{Concept}}, {{Modeling}}, and {{Analysis}}},
	shorttitle = {Harmonic {{Stability}} in {{Power Electronic-Based Power Systems}}},
	author = {Wang, Xiongfei and Blaabjerg, Frede},
	year = {2019},
	month = may,
	journal = {IEEE Transactions on Smart Grid},
	volume = {10},
	number = {3},
	pages = {2858--2870},
	issn = {1949-3061},
	doi = {10.1109/TSG.2018.2812712},
	urldate = {2024-11-10}
}

@inproceedings{kamalinejadImpactControlParameters2025,
  title = {Impact of {{Control Parameters}} on {{Angular Stability}} of {{Grid-Forming Converters Using Virtual-Admittance Based Control}}},
  booktitle = {2025 {{Energy Conversion Congress}} \& {{Expo Europe}} ({{ECCE Europe}})},
  author = {Kamalinejad, Kavian and Narula, Anant and Bongiorno, Massimo and Beza, Mebtu and Svensson, Jan R.},
  year = {2025},
  month = sep,
  pages = {1--6},
  doi = {10.1109/ECCE-Europe62795.2025.11238912},
  urldate = {2026-02-04}
}

@article{imgartDecoupledPQGridForming2026,
  title = {Decoupled {{PQ Grid-Forming Control With Tunable Converter Frequency Behaviour}}},
  author = {Imgart, Paul and Narula, Anant and Bongiorno, Massimo and Beza, Mebtu and Svensson, Jan R. and Hasler, Jean-Philippe and Mattavelli, Paolo},
  year = {2026},
  journal = {IEEE Open Journal of Industry Applications},
  pages = {1--15},
  issn = {2644-1241},
  doi = {10.1109/OJIA.2026.3658761},
  urldate = {2026-02-04}
}

\end{document}